\documentclass[a4paper,onecolumn,fleqn,accepted=2025-12-23]{quantumarticle}
\pdfoutput=1

\usepackage[utf8]{inputenc}
\usepackage[T1]{fontenc}
\usepackage{microtype}

\usepackage{csquotes}
\usepackage[english]{babel}

\usepackage{graphicx}
\usepackage[dvipsnames]{xcolor}
\usepackage[colorlinks=true, breaklinks=true]{hyperref}
\usepackage[hyphenbreaks]{breakurl}

\usepackage{mathtools, amssymb, bm, mathrsfs}
\usepackage{tensor, braket}

\usepackage[capitalise]{cleveref}

\usepackage{aas_macros}
\usepackage[
	sorting=none,
	style=numeric-comp,
	giveninits=true,
	doi=true,
	backref=false,
	date=year,
	urldate=iso,
    seconds=true,
    eprint=true,
    isbn=false
	]{biblatex}
\hypersetup{breaklinks=true}
\AtEveryBibitem{\clearfield{labeldate}}
\AtEveryBibitem{\clearfield{pagetotal}}
\AtEveryBibitem{\iffieldundef{eid}{}{\clearfield{pages}}}
\AtEveryBibitem{%
  \ifentrytype{article}
    {\clearfield{eprint}}
    {}}
\AtEveryBibitem{\clearfield{eprintclass}}

\newbibmacro{string+doi}[1]{\iffieldundef{doi}{#1}{\href{https://doi.org/\thefield{doi}}{#1}}}
\DeclareFieldFormat{title}{\usebibmacro{string+doi}{\mkbibemph{#1}}}
\DeclareFieldFormat[article]{title}{\usebibmacro{string+doi}{\mkbibquote{#1}}}
\DeclareFieldFormat[incollection]{title}{\usebibmacro{string+doi}{\mkbibquote{#1}}}
\DeclareFieldFormat[unpublished]{title}{\usebibmacro{string+doi}{\mkbibquote{#1}}}

\numberwithin{equation}{section}

\DeclareMathOperator{\re}{Re}
\DeclareMathOperator{\rank}{rank}
\DeclareMathOperator{\kernel}{ker}
\DeclareMathOperator{\image}{im}
\let\t\tensor
\let\p\partial

\def\ee{\mathrm e}
\def\ii{\mathrm i}
\def\p{\partial}
\def\half{\tfrac{1}{2}}
\def\ihalf{\tfrac{\ii}{2}}

\def\mfd{M}
\def\sflds{\mathscr C}
\def\vflds{\mathfrak X}

\def\Sh{S}
\def\Sd{S^\dagger}
\def\del{{\mathring \nabla}}
\def\dd{\mathrm{d}}
\def\D{D}
\def\Dp{D'}
\def\Dd{{\mathscr D}}

\def\numR{\mathbb{R}}
\def\numC{\mathbb{C}}

\newcommand{\ketbra}[2]{\ket{#1}\!\!\bra{#2}}

\def\berA{\mathscr A}
\def\berF{\mathscr F}

\def\Dir{\boldsymbol D}

\def\spinorsB{\mathscr S}   
\def\spinorsP{\varpi}       

\begin{document}
\title{Quantum geometric tensors from sub-bundle geometry}

\author{Marius A. Oancea}
\orcid{0000-0002-1242-4041}
\email{marius.oancea@univie.ac.at}
\affiliation{University of Vienna, Faculty of Physics, Boltzmanngasse~5, 1090 Vienna, Austria}

\author{Thomas B. Mieling}
\orcid{0000-0002-6905-0183}
\email{thomas.mieling@univie.ac.at}
\affiliation{University of Vienna, Faculty of Physics and Research Network TURIS, Boltzmanngasse~5, 1090 Vienna, Austria}

\author{Giandomenico Palumbo}
\orcid{0000-0003-1303-1247}
\email{giandomenico.palumbo@gmail.com}
\affiliation{School of Theoretical Physics, Dublin Institute for Advanced Studies, 10 Burlington Road, Dublin 4, Ireland}

\maketitle

\begin{abstract}
The geometric properties of quantum states are crucial for understanding many physical phenomena in quantum mechanics, condensed matter physics, and optics. The central object describing these properties is the quantum geometric tensor, which unifies the Berry curvature and the quantum metric. In this work, we use the differential-geometric framework of vector bundles to analyze the properties of parameter-dependent quantum states and generalize the quantum geometric tensor to this setting. This construction is based on a general connection on a Hermitian vector bundle, which defines a notion of quantum state transport in parameter space, and a sub-bundle projector, which constrains the set of accessible quantum states. We show that the sub-bundle geometry is similar to that of submanifolds in Riemannian geometry and is described by generalized Gauss–Codazzi–Mainardi equations. This leads to a novel definition of the quantum geometric tensor that contains an additional curvature contribution. To illustrate our results, we describe the sub-bundle geometry arising in the semiclassical treatment of Dirac fields propagating in curved spacetime and show how the quantum geometric tensor, with its additional curvature contributions, is obtained in this case. As a concrete example, we consider Dirac fermions confined to a hyperbolic plane and demonstrate how spatial curvature influences the quantum geometry. This work sets the stage for further exploration of quantum systems in curved geometries, with applications in both high-energy physics and condensed matter systems.
\end{abstract}

\section{Introduction}

Geometric phases are fundamental concepts in quantum mechanical systems, describing phase differences acquired by the wave functions of quantum states that are induced by the non-trivial geometry of the parameter space in which the systems evolve \cite{Berry,Geometric_phases_book}. Importantly, because of their geometric origin \cite{PhysRevLett.51.2167,Geometric_phases_book}, they are independent of the specific dynamical evolution of the quantum states.

Quantum geometry has emerged as a central concept in modern condensed matter physics \cite{Resta,Torma,Xie,Torma2,Holder1}, providing insights into various quantum phenomena through the quantum geometric tensor. This complex rank-2 tensor is a gauge-invariant physical quantity that has already been measured in several experimental setups \cite{Plenio,Sanvitto,Cappellaro,Comin}.
Moreover, this tensor unifies the Berry curvature \cite{Berry,PhysRevLett.51.2167} and the quantum metric \cite{Provost1980} that have been used to characterize the topology and geometry of quantum states in parameter spaces \cite{Zanardi,Gritsev,Levay,Vergara}, momentum spaces \cite{Malpuech,Ryu,Palumbo1,Ahn,Northe,Piechon,Bouhon,Holder3}, and configuration spaces \cite{Marrazzo,Chen2}. Recent developments have significantly broadened the scope of quantum geometry, extending its range of applicability to include degenerate quantum states \cite{PhysRevB.81.245129,Neupert, Mera2,Ding, PhysRevB.107.245136,avdoshkin2024geometry,Hosur,Jankowski2,mitscherling2024gauge}, flat-band superconductors \cite{Peotta1,Peotta2,Bernevig,BoYang}, tensor monopoles \cite{Palumbo2018,Ding2,Tan}, Euler insulators \cite{Jankowski,Jankowski3,Kwon}, non-Hermitian physics \cite{PhysRevA.99.042104,Zhu,Ilan,Strasser,Smith}, out-of-equilibrium systems \cite{Salerno,Hamma,Zhou2024,Pandey}, interacting phases \cite{Roy,Thomale,PhysRevB.104.045104,JieWang,
JieWang2,Salerno2,Chen,Sukhachov,Bergholtz,Holder2,Parker,Fujimoto,Tesfaye}, quantum systems with spin-topology \cite{Lange}, geometric semimetals \cite{Lin2}, finite-temperature mixed states \cite{Mera3,Chien}, and more. These studies have underscored the importance of generalizing the notion of quantum metrics to accommodate more intricate scenarios encountered in contemporary research.

In this paper, we present a general construction of the quantum geometric tensor within the differential geometric framework of vector bundles \cite{1987_Kobayashi,2012_Demailly}. This can be related to the usual descriptions of quantum geometry in condensed matter physics by identifying the base space of the vector bundle with the parameter space of a quantum system and the fibers of the vector bundle with finite-dimensional Hilbert spaces, thus describing the parameter-dependent space of quantum states. To cover as many physical scenarios as possible, we only introduce a minimum amount of geometric structure, and all our calculations are presented in a coordinate-independent way. More concretely, we assume the complex vector bundles to be equipped with a Hermitian pseudo-metric, a compatible connection, and an orthogonal projector, which defines an orthogonal decomposition of the bundle. The sub-bundle geometry induced by the projector and the connection represents the quantum geometry usually studied in condensed matter physics and quantum mechanics \cite{Provost1980,PhysRevB.81.245129,PhysRevB.107.245136}.
This geometric framework allows for natural definitions of the quantum geometric tensor, the Berry connection, and the quantum metric, which generalize previous constructions in several ways.

From a mathematical perspective, the induced sub-bundle geometry is very similar to the classical problem in Riemannian geometry regarding the geometry of hypersurfaces \cite{1999_Spivak_3}, or more generally, the geometry of submanifolds \cite{1999_Spivak_4,2019_Dajczer}. We show that the sub-bundle geometry is described by a generalized form of the Gauss–Codazzi–Mainardi equations (similar generalizations are also given in Refs.~\cite{1987_Kobayashi,2012_Demailly}, although only for the case of holomorphic vector bundles over complex manifolds). The main geometric objects that describe the sub-bundle geometry are the projected connection, which is essentially the Berry connection, the first fundamental form represented by the fiber metric, and the second fundamental form or shape operator \cite[p.~235]{2018_Lee}. The quantum geometric tensor then arises as a derived object that is obtained by contracting the shape operator with its adjoint via the first fundamental form, which can be seen as a generalization of the third fundamental form of classical differential geometry \cite[p.~62]{1999_Spivak_3}.

The first mathematical definition of the quantum metric was given in Ref.~\cite{Provost1980} as a measure of the distance between neighboring quantum states that depend on a set of parameters. The quantum geometric tensor, which unifies the quantum metric and the Berry curvature, was later introduced by Berry \cite{berry1989QGT} (see also Ref.~\cite{Geometric_phases_book} for a textbook treatment). More general definitions of the quantum geometric tensor in Hermitian systems, covering the non-Abelian case and emphasizing the role of the projector, were given later in Refs.~\cite{PhysRevB.81.245129,PhysRevB.107.245136,avdoshkin2024geometry}. Compared with these previous works, our results provide a more general definition of the quantum geometric tensor in the following ways. First, some of the previous works assume that the parameter space is $\numR^N$, while in our case it is represented by a generic manifold. Second, and most importantly, in all previous papers a flat connection was assumed when taking derivatives of quantum states, and the Berry connection was defined by projecting the flat connection onto the relevant subspace. Working with a general connection on the complex vector bundle instead, we obtain an additional curvature contribution to the quantum geometric tensor. This shows that the common statement that “the antisymmetric part of the quantum geometric tensor gives the Berry curvature” is only true when the Berry connection is the projection of a flat connection. Similarly to Ref.~\cite{PhysRevB.81.245129}, our treatment of the quantum geometric tensor covers both the Abelian and non-Abelian cases, since we do not make any particular assumptions about the rank of the projector, besides being constant. Another general feature of our approach is that we work with a general Hermitian pseudo-metric which is allowed to vary between different points in the base manifold.

A more general notion of the quantum geometric tensor, especially by allowing the parameter space to be a manifold and working with a general connection, is fundamental for the proper description of certain physical scenarios. For example, the Berry curvature is essential for the description of semiclassical wave packet dynamics \cite{Niu,Niu2}, but general-relativistic extensions of these results require working on manifolds and using non-flat connections. Such generalized forms of Berry curvature have been shown, for instance, to determine gravitational spin Hall effects \cite{GSHE_reviewCQG}, which govern the spin-dependent semiclassical propagation of electromagnetic \cite{SHE_QM1,GSHE2020,Harte_2022,PhysRevD.109.064020}, linearized gravitational \cite{SHE_GW,GSHE_GW,GSHE_lensing,GSHE_lensing2}, and Dirac \cite{SHE_Dirac,PhysRevD.107.044029} wave packets in curved spacetime. As an application of our geometric framework, we show how the quantum geometric tensor can be obtained for semiclassical Dirac fields on general spacetimes. In this way, we recover the Berry curvature introduced in Ref.~\cite{PhysRevD.107.044029}, and we also obtain the corresponding quantum metric, which naturally encodes unique aspects of quantum states coupled to a non-trivial spacetime geometry. In fact, in addition to generalizing the static phase-space Berry curvature discussed in Refs.~\cite{Niu,Niu2,Hayata} and the more recent time-dependent quantum geometric tensor presented in Refs.~\cite{Hamma,Queiroz,Vergara2}, our formulation of the non-Abelian quantum metric induced by a spacetime metric is completely novel.

This paper is structured as follows. \Cref{s:geometric setup} presents the general differential geometric setup. We review the standard definitions of connections and curvature in complex vector bundles and introduce the orthogonal projectors. In \cref{s:projected connections}, we present our general theory for the quantum geometric tensor, based on the central role of projected covariant derivatives and shape operators. Finally, in \cref{s:applications}, we discuss some applications of our results in the context of Dirac fermions on curved spacetime. In particular, to provide a more direct application of our framework in condensed matter physics, we use our generalized quantum metric to study Dirac fermions on the hyperbolic plane, which are relevant in the context of topological phases on hyperbolic lattices \cite{Kollar,Yu,Zhang2,Bzdusek,Bzdusek2,Bzdusek3} in the low-energy regime. Our results are not only relevant for understanding and investigating the quantum geometry of hyperbolic topological matter, but provide a robust theoretical foundation for future explorations of quantum systems coupled to static and dynamical background geometries across solid-state and synthetic matter systems of any dimension \cite{Celi,Ojanen,Marzuoli,Lambiase,Segev,Para,Palumbo2017,Pachos,Haller,Wagner,Nissinen,Palumbo2016,Gorini}.

\section{Differential geometric setup}
\label{s:geometric setup}

Our main goal is the description of parameter-dependent quantum systems and their evolution under slow changes in parameter space.
This means that for each parameter $\lambda$ in a parameter space $M$, there is a finite-dimensional space of states $\mathscr H(\lambda)$ that is isomorphic (though not necessarily naturally) to $\numC^n$, with the dimension $n$ being the same for all values of $\lambda$.
Formally, such a structure is a vector bundle $E$ over $M$ with typical fiber $\numC^n$.
In this picture, a constraint (that may depend on the parameters $\lambda$) implies that not all states in $E$ are accessible. Instead, if the number of constraints is constant for all $\lambda$, the accessible states form a sub-bundle $E'$ of $E$ (with typical fiber $\numC^m$ where $m < n$).
Such descriptions apply, for example, to condensed matter systems, where electrons in periodic lattices are constrained to certain energy bands \cite{Mermin}, or to Wentzel–Kramers–Brillouin (WKB)-type approximations, where the amplitude is constrained to the eigenspace of the principal symbol of the corresponding differential operator \cite{PhysRevA.44.5239,Emmrich1996,PhysRevD.107.044029}.

The theory developed here requires further mathematical structures on the bundle $E$, namely a (potentially indefinite) fiber metric and a compatible connection.
If all spaces $\mathscr H(\lambda)$ are Hilbert spaces, then the various inner products define a fiber metric (provided that they vary continuously with $\lambda$). However, we do not restrict to such positive-definite products, but also allow for an indefinite product as arising in the case of Dirac $4$-spinors or for pseudo-Hermitian quantum systems \cite{Mosta,Mosta2}, for example.
In either case, in order to compare states in different state spaces, a connection on $E$ is required.

The aim of this section is to make this heuristic picture mathematically precise. More details on the differential geometry of complex vector bundles can be found in Refs.~\cite{1987_Kobayashi,2012_Demailly}.

\subsection{Bundles, connections, and curvature}

Let $\mfd$ be a real $C^\infty$ manifold of dimension $N < \infty$ and denote by $\sflds_\numR(\mfd)$ and $\sflds_\numC(\mfd)$ the rings of real and complex scalar functions on $\mfd$, respectively, and by $\vflds(\mfd)$ the module of real vector fields on $\mfd$ (all assumed to be smooth).
Depending on the physical setting, the manifold $\mfd$ can represent different objects, such as spacetime, a Lagrangian submanifold (in the context of WKB-type semiclassical treatments \cite{Emmrich1996, PhysRevD.107.044029}), momentum space (in condensed matter scenarios \cite{Malpuech,Ryu,Palumbo1,Ahn,Northe,Piechon,Bouhon,Holder3}), or some general parameter space \cite{Provost1980,Gritsev,Levay,Vergara,PhysRevB.81.245129,PhysRevB.107.245136}. However, in most previous treatments of the quantum geometric tensor, the parameter space was assumed to be $\numR^N$.

We consider a complex vector bundle $(E, \mfd, \pi_E, \numC^n)$ with base space $\mfd$, projection ${\pi_E: E \to \mfd}$, and typical fiber $\numC^n$, but all the arguments below can be carried over to the real setting without significant modification.
In either case, the space of sections of $E$ will be denoted by $\Gamma(E)$.

A non-degenerate Hermitian form on $E$ is a map $h: \Gamma(E) \times \Gamma(E) \to \sflds_\numC(\mfd)$ satisfying
\begin{subequations}
\begin{align}
    &h(\Psi, \cdot) = 0 \Leftrightarrow \Psi = 0\,,
    \\
    &h(\Psi_1, \Psi_2) = \overline{h(\Psi_2, \Psi_1)}\,,
    \\
    &h(\Psi_1, \Psi_2 + f \Psi_3) = h(\Psi_1, \Psi_2) + f h(\Psi_1, \Psi_3)\,,
\end{align}
\end{subequations}
where $\Psi, \Psi_1, \Psi_2, \Psi_3 \in \Gamma(E)$ are arbitrary sections, $f \in \sflds_\numC(\mfd)$ is any complex scalar field, and overlines denote complex conjugation. With this additional structure, the bundles are referred to as Hermitian vector bundles \cite{1987_Kobayashi,2012_Demailly}.
Note that this does not entail that $h$ is a Hermitian \emph{metric} on the fibers of $E$ since positive-definiteness is not assumed—instead $h$ can be regarded as a Hermitian pseudo-metric on the fibers.
The action of the form $h$ can also be translated into the Dirac notation as $h(\Phi, \Psi) = \braket{\Phi \mid \Psi}$.

A covariant derivative (also referred to as a connection) on $E$ is a map
\begin{align}
    & \nabla: \vflds(\mfd) \times \Gamma(E) \to \Gamma(E)
    &
    & (X, \Phi) \mapsto \nabla_X \Phi
\end{align}
such that for all vector fields $X, X_1, X_2 \in \vflds(\mfd)$, sections $\Phi, \Phi_1, \Phi_2 \in \Gamma(E)$, real-valued scalar fields $\lambda \in \sflds_\numR(\mfd)$ and complex scalar fields $f \in \sflds_\numC(\mfd)$ one has
\begin{subequations}
\begin{align}
    \nabla_{X_1 + \lambda X_2} \Phi
        &= \nabla_{X_1} \Phi + \lambda \nabla_{X_2} \Phi
        \,,
    \\
    \nabla_X(\Phi_1 + f\, \Phi_2)
        &= \nabla_X \Phi_1 + f \nabla_X \Phi_2 + (X f) \Phi_2
        \,,
\end{align}
\end{subequations}
where $Xf$ denotes the directional derivative of $f$ along $X$.
This connection on $E$ naturally induces further connections on $E'$, the dual bundle of $E$, as well as their tensor products. The connection $\nabla$ is said to be compatible with the Hermitian form $h$ if, for all $X \in \vflds(\mfd)$ and $\Phi, \Psi \in \Gamma(E)$, one has
\begin{align}
    \label{eq:derivative h compatibility}
    X[h(\Phi, \Psi)]
        &= h(\nabla_X \Phi, \Psi) + h(\Phi, \nabla_X \Psi)\,.
\end{align}
In the following, this compatibility condition will be assumed throughout.

The curvature of the covariant derivative $\nabla$ is the unique multilinear map
\begin{align}
    &\Omega : \vflds(\mfd) \times \vflds(\mfd) \times \Gamma(E) \to \Gamma(E)
    &
    &(X, Y, \Phi) \mapsto \Omega(X, Y) \Phi\,,
\end{align}
that satisfies
\begin{align} \label{eq:Ricci identity}
    \nabla_X \nabla_Y \Phi - \nabla_Y \nabla_X \Phi - \nabla_{[X, Y]} \Phi
        = \Omega(X, Y) \Phi\,,
\end{align}
where $[X, Y]$ denotes the Lie bracket of $X$ and $Y$.
The curvature satisfies $\Omega(X, Y) = - \Omega(Y, X)$ and (due to the compatibility of $\nabla$ with $h$)
\begin{align}
    h[\Omega(X, Y) \Phi, \Psi]
        = - h[\Phi, \Omega(X, Y) \Psi]\,.
\end{align}
We emphasize that this geometric setup is more general than that of previous treatments of the quantum geometric tensor, where the parameter space was generally assumed to be $\numR^N$ and a flat connection $\nabla = \partial$ with $\Omega = 0$ was used.
Moreover, while our analysis allows for general $\nabla$, the specific connection in concrete applications cannot be chosen arbitrarily, but is determined by the physical problem.
For example, the relevant connection entering the semiclassical dynamics of spinors in curved spacetimes or hyperbolic Dirac materials is determined by the Levi-Civita connection and the spin structure of the spacetime manifold, as discussed in \cref{s:applications} below.

\subsection{Frames and index notation}
\label{s:frames and indices}

A local frame of $E$ over an open subset $U \subseteq M$ consists of an ordered collection of local sections $(s_i) = (s_1, \ldots, s_n)$ with $s_i \in \Gamma(E|_U)$ such that $\{ s_1(x), s_2(x), \ldots , s_n(x) \}$ forms a basis of the fiber $E_x$ for all $x \in U$.
Given such a local frame, the associated coframe $(\sigma^i)$ consists, at each point, of the dual basis vectors that satisfy $\t\sigma{^i}(\t s{_j}) = \t*\delta{^i_j}$.
Relative to a frame $(s_i)$, $h$ is described by the fields
\begin{align}
    \t h{_i_j}
        = h(s_i, s_j)\,,
\end{align}
which assign to each point $x \in U$ a non-degenerate Hermitian matrix $(\t h{_i_j})$.
The connection $\nabla$, on the other hand, is described by a collection of one-forms acting on a vector field $X$ as
\begin{align}
    \t\omega{^i_j}(X)
        = \t\sigma{^i}(\nabla_X \t s{_j})\,.
\end{align}
Since $h$ is assumed to be non-degenerate, $\t\omega{^i_j}$ is uniquely characterized by the quantities
\begin{align}
    \t\omega{_i_j}(X)
        = \t h{_i_k} \t\omega{^k_j}(X)
        = h(\t s{_i}, \nabla_X \t s{_j})\,.
\end{align}
For the following analysis, it is beneficial to work with a frame relative to which $\t h{_i_j}$ is a constant matrix (typical applications reduce $\t h{_i_j}$ to a “standard form” in which all diagonal components of $\t h{_i_j}$ are $+1$ or $-1$, and all off-diagonal entries vanish). In this case, the compatibility condition \eqref{eq:derivative h compatibility} takes the form
\begin{align}
    \label{eq:derivative h compatibility indices}
    \t\omega{_i_j} + \t{\bar\omega}{_j_i} = 0\,,
\end{align}
where we use the notation $\t{\bar\omega}{_j_i} = \overline{\t{\omega}{_j_i}}$ for the complex conjugation of tensor components.
In terms of these connection coefficients, the curvature $\Omega$ can be expressed as
\begin{align}
    \t\Omega{^i_j}(X,Y)&
        = X[\t\omega{^i_j}(Y)] - Y[\t\omega{^i_j}(X)] - \t\omega{^i_j}([X,Y])
        + \t\omega{^i_k}(X) \t\omega{^k_j}(Y) - \t\omega{^i_k}(Y) \t\omega{^k_j}(X)\,.
\end{align}
Regarding $\t\omega{^i_j}$ and $\t\Omega{^i_j}$ as matrix-valued differential forms, this can be written as
\begin{align}
    \t\Omega{^i_j}
        = \dd \t\omega{^i_j}
        + \t\omega{^i_k} \wedge \t\omega{^k_j}\,,
\end{align}
where $\dd$ and $\wedge$ denote exterior derivatives and exterior products, respectively.
In particular, relative to a coordinate frame, the above formula reduces to
\begin{align}
    \t\Omega{^i_j_\mu_\nu}
        &= \t\p{_\mu} \t\omega{^i_j_\nu} - \t\p{_\nu} \t\omega{^i_j_\mu}
        + \t\omega{^i_k_\mu} \t\omega{^k_j_\nu} - \t\omega{^i_k_\nu} \t\omega{^k_j_\mu}\,.
\end{align}

\subsection{Projections and sub-bundles}
\label{s:projectors}

If $E$ is to be interpreted as an unconstrained space of states, imposing a set of linear constraints can be formulated in terms of a projector $P$.
This is an $h$-compatible bundle-endomorphism of $E$ satisfying
\begin{align}
    P^2 &= P\,,
    &
    h(\Phi, P \Psi) = h(P \Phi, \Psi)\,.
\end{align}
The intuition behind this notion is that the “space of constrained states” $E_\parallel = \image P$ contains all states satisfying a system of linear equations $P \Phi = \Phi$, while the complement $E_\perp = \kernel P$ contains all remaining states satisfying $P \Psi = 0$. The sub-bundles $E_\parallel = \image P$ and $E_\perp = \kernel P$ then yield the orthogonal decomposition
\begin{align}
    E = E_\parallel \oplus E_\perp\,.
\end{align}
Given a projector $P$, it is beneficial to use adapted frames of the form $(\t s{_i}) = (\t s{_A}, \t s{_I})$, where the sections $\t s{_A}$ and $\t s{_I}$ span the fibers of $E_\parallel$ and $E_\perp$, respectively.
The corresponding dual frame $(\t \sigma{^i}) = (\t \sigma{^A}, \t \sigma{^I})$ then satisfies
\begin{align}
    \t \sigma{^A}(\t s{_B}) &= \t*\delta{^A_B}\,,
    &
    \t \sigma{^A}(\t s{_I}) &= 0\,,
    &
    \t \sigma{^I}(\t s{_B}) &= 0\,,
    &
    \t \sigma{^I}(\t s{_J}) &= \t*\delta{^I_J}\,.
\end{align}
The mathematical picture of these projectors and the associated sub-bundle geometry is realized, for example, in condensed matter physics when electrons are constrained to certain energy bands \cite{Mermin}. Another example of such a decomposition arises when considering WKB-type approximations for multicomponent fields, where the WKB amplitude is constrained to an eigenspace of the principal symbol of the corresponding differential operator \cite{PhysRevA.44.5239,Emmrich1996,PhysRevD.107.044029}. The rank of $P$ is assumed to be constant on the parameter manifold $M$, and similarly to the analysis of Ref.~\cite{PhysRevB.81.245129} this can lead to an Abelian ($\rank P = 1$) or non-Abelian ($1 < \rank P < n$) quantum geometric tensor.

The projector $P$ is compatible with the Hermitian form $h$, but need not be compatible with the covariant derivative $\nabla$.
The consequences of $\nabla_X P \neq 0$, described in detail in \cref{s:projected connections} below, are the main source of the interesting properties associated with Berry connections, quantum geometric tensors, and quantum geometry in general.

\section{Berry connection, shape operators, and derived quantities}
\label{s:projected connections}

In this section, we present our main mathematical results regarding the quantum geometry, or in other words the sub-bundle geometry, determined by the covariant derivative $\nabla$ and the projector $P$.
In particular, we identify the shape operators as central geometric quantities as they fully determine the quantum geometric tensor, and provide a geometric decomposition of the quantum geometric tensor into \emph{three} parts, namely the quantum metric, the Berry curvature, and the bundle curvature (which was absent in previous work on the subject).

\subsection{Berry connection and shape operators}

We start here by introducing the main geometric objects that describe the sub-bundle geometry. These are the projected covariant derivatives, which are directly related to the Berry connection, and the shape operators or the second fundamental forms, which describe the difference between $\nabla$ and the projected connections via the Weingarten equations \cite[p.~35]{1999_Spivak_4}.

\subsubsection{Projected covariant derivatives}
In general, the covariant derivative $\nabla$ is not compatible with the projection $P$. This means that if $\Phi_\parallel$ and $\Psi_\perp$ are sections of $E$ taking values in $E_\parallel$ and $E_\perp$, respectively, then $\nabla_X \Phi_\parallel$ and $\nabla_X \Psi_\perp$ are not guaranteed to take values in $E_\parallel$ and $E_\perp$, respectively.
This can be seen as follows. Writing $P' = I - P$ for the projection onto $E_\perp$, one has
\begin{align}
    \label{eq:derivative decomposition}
    \nabla_X \Phi_\parallel &
        = P \nabla_X \Phi_\parallel
        + P'\nabla_X \Phi_\parallel\,,
    &
    \nabla_X \Psi_\perp &
        = P' \nabla_X \Psi_\perp
        + P \nabla_X \Psi_\perp\,.
\end{align}
The second terms in both equations, describing the contributions in the respective complementary bundles, can be expressed in terms of $\nabla_X P$, namely
\begin{align}
    \label{eq:complementary projections of nabla}
    P' \nabla_X \Phi_\parallel
        &= + (\nabla_X P) \Phi_\parallel\,,
    &
    P \nabla_X \Psi_\perp
        &= - (\nabla_X P) \Psi_\perp\,.
\end{align}
This shows that taking $\nabla$-derivatives of sections of $E_\perp$ or $E_\parallel$ leads out of these bundles whenever $\nabla_X P \neq 0$.
This can be resolved by defining the projected connections:
\begin{align}
    \D{_X} \Phi_\parallel
        &= P(\t\nabla{_X} \Phi_\parallel)\,,
    &
    \Dp_X \Psi_\perp
        &= P'(\t\nabla{_X} \Psi_\parallel)\,.
\end{align}

Using the methods described in \cref{s:frames and indices}, we can compute the connection coefficients and curvature tensors for these two covariant derivatives. In particular, the Berry connection coefficients are \cite{Geometric_phases_book}
\begin{align}
    \t\berA{_A_B_\mu}
        = \ii \t*\omega{^\parallel_A_B_\mu}
        = \ii h( \t s{_A} , \t\D{_\mu} \t s{_B} )
        = \ii h( \t s{_A} , \t\nabla{_\mu} \t s{_B} )
        \,,
\end{align}
which are Hermitian in $A$ and $B$, i.e., $\t\berA{_A_B_\mu} = \t{\bar\berA}{_B_A_\mu}$. We can also write the Berry connection coefficients in Dirac notation as $\t\berA{_A_B_\mu} = \ii \braket{ \t s{_A} \mid \t\nabla{_\mu} \t s{_B} }$.
In terms of these coefficients, the Berry curvature takes the form
\begin{align}
    \t\berF{_A_B_\mu_\nu}
        = \ii \t*\Omega{^\parallel_A_B_\mu_\nu}
        = \t\p{_\mu} \t\berA{_A_B_\nu} - \t\p{_\nu} \t\berA{_A_B_\mu}
        - \ii \t h{^C^D} (\t\berA{_A_C_\mu} \t\berA{_D_B_\nu} - \t\berA{_A_C_\nu} \t\berA{_D_B_\mu})\,,
\end{align}
where $\t h{^C^D}$ is the inverse of $\t h{_A_B}$.

\subsubsection{Shape operators}

\Cref{eq:complementary projections of nabla} shows that the differences between $\nabla$ and the projected connections $\D$ and $\D'$ can be understood in terms of the shape operators
\begin{subequations}
\begin{align}
    \label{def:shape operator 1}
    \Sh : \Gamma(E_\parallel) \times \vflds(M) &\to \Gamma(E_\perp)\,,
    &
    S(\Phi_\parallel, X) &= (\nabla_X P) \Phi_\parallel\,,
    \\
    \label{def:shape operator 2}
    \Sd : \Gamma(E_\perp) \times \vflds(M) &\to \Gamma(E_\parallel)\,,
    &
    \Sd(\Phi_\perp, X) &= (\nabla_X P) \Phi_\perp\,.
\end{align}
\end{subequations}
The notation is justified by the fact that the operators $\Sh_X \equiv \Sh(\cdot, X)$ and $\Sd_X \equiv \Sd(\cdot, X)$ are adjoint in the sense that
$h[\Phi_\perp, \Sh^{\phantom\dagger}_X \Psi_\parallel] = h[\Sd_X \Phi_\perp, \Psi_\parallel]$,
which can be shown by differentiating $h(\Phi_\perp, \Psi_\parallel) = 0$.
The expression $h[\Phi_\perp, \Sh^{\phantom\dagger}_X \Psi_\parallel] $ arising here is multilinear in the arguments $\Phi_\perp$, $\Psi_\parallel$ and $X$, and is known as the second fundamental form:
\begin{align}
    & \underline S : \Gamma(E_\perp) \times \Gamma(E_\parallel) \times \vflds(M) \to \sflds_\numC\,,
    &
    & \underline S(\Phi_\perp, \Psi_\parallel; X) = h[\Phi_\perp, \Sh_X \Psi_\perp]\,.
\end{align}
Combining \cref{eq:derivative decomposition,eq:complementary projections of nabla} leads to the Weingarten equations
\begin{align}
    \label{eq:Weingarten}
    \nabla_X \Phi_\parallel
        &= \D_X \Phi_\parallel + \Sh^{\phantom\dagger}_X(\Phi_\parallel)\,,
    &
    \nabla_X \Psi_\perp
        &= \Dp_X \Psi_\perp - \Sd_X(\Psi_\perp)\,.
\end{align}

Relative to an adapted frame, the components of the shape operators are given by
\begin{align}
    \t S{^I_A_\mu} &= \t \sigma{^I}[(\t\nabla{_\mu} P) \t s{_A}]\,,
    &
    \t S{^A_I_\mu} &= \t \sigma{^A}[(\t\nabla{_\mu} P) \t s{_I}]\,,
\end{align}
such that
\begin{align}
    \Sh(\Phi, X) &= \t S{^I_A_\mu} \t \Phi{^A} \t X{^\mu} \t s{_I}\,,
    &
    \Sd(\Psi, X) &= \t S{^A_I_\mu} \t \Psi{^I} \t X{^\mu} \t s{_A}\,.
\end{align}
Since $\t h{_i_j}$ is non-degenerate, the components $\t S{^I_A_\mu}$ are uniquely determined by those of the second fundamental form
$\t S{_I_A_\mu} = h[\t s{_I}, (\t\nabla{_\mu} P) \t s{_A}] = h(\t s{_I}, \t\nabla{_\mu} \t s{_A})$.
In Dirac notation, these coefficients can be written as $\t\Sh{_I_A_\mu} = \braket{\t s{_I} \mid \t{\nabla\!}{_\mu} \t s{_A}}$, which shows that the three-index “$N$-beins” of Refs.~\cite{Romero_2024,Northe,PhysRevB.89.235424} and the “quantum Christoffel symbols of the first kind” introduced in Ref.~\cite{2023JPhA...56T5301R} can be interpreted as frame components of the second fundamental form $\underline S$ (the latter quantity should not be confused with the Christoffel symbols or connection coefficients derived from the quantum metric described in \cref{s:quantum metric} below \cite{Romero_2024,mitscherling2024gauge}).

In the current geometric setup, the covariant derivative of the shape operator is undefined.
This is because the index $\mu$ in the expressions $\t S{^I_A_\mu}$ and $\t S{^A_I_\mu}$ refers to $T\mfd$, the tangent bundle of $M$, and the only derivatives considered so far are defined in $E$ and its sub-bundles $E_\parallel$ and $E_\perp$.
To differentiate such objects, a covariant derivative $\del$ is required on the base manifold $\mfd$. Then, we can define a covariant derivative on the tensor products of $E_\parallel$, $E_\perp$, and $T M$ (as well as their duals) via the rule
\begin{align}
    \Dd_X (\Phi_\parallel \otimes \Psi_\perp \otimes Y)
        = (\D_X \Phi_\parallel) \otimes \Psi_\perp \otimes Y
        + \Phi_\parallel \otimes (\Dp_X \Psi_\perp) \otimes Y
        + \Phi_\parallel \otimes \Psi_\perp \otimes ( \del_X Y)\,.
\end{align}
For example, for the shape operator $\Sh$ one then has
\begin{align}
    \label{eq:derivative of shape operator}
    \Dp_X [\Sh(\Phi_\parallel, Y)]
        = (\Dd_X \Sh)(\Phi_\parallel, Y)
        + \Sh(\D_X \Phi_\parallel, Y)
        + \Sh(\Phi_\parallel, \del_X Y)\,.
\end{align}
The following analysis will make use of such a derivative $\del$ for intermediate calculations – the final results, however, will be independent of this connection.

\subsection{Berry curvature and the Gauss–Codazzi–Mainardi equations}

Since the Berry connection is derived from the covariant derivative $\nabla$ and the projection $P$, the Berry curvature of the projected connection can be expressed in terms of the projection of the curvature $\Omega$ and the shape operators $\Sh$ and $\Sd$.
The precise relation is given by generalizations of the Gauss equation known from Riemannian (or pseudo-Riemannian) geometry. We also derive generalizations of the Codazzi–Mainardi equations that relate the covariant derivatives of the shape operators to the “mixed components” of $\Omega$ with respect to the projections $P$ and $P'$. A similar generalization of the Gauss–Codazzi–Mainardi equations can also be found in Refs.~\cite{1987_Kobayashi,2012_Demailly}, although only for the case of holomorphic vector bundles over complex manifolds. In Ref.~\cite{PhysRevB.107.245136}, submanifold geometry and the Gauss–Codazzi–Mainardi equations are also used in connection with the quantum geometric tensor, but in a different way compared to our current treatment and for the specific case of submanifolds in $\numC P^n$.

\subsubsection{Gauss equations}
The Gauss equations show that curvatures $\Omega^\parallel$ (of $D$) and $\Omega^\perp$ (of $D'$) are fully determined by the curvature $\Omega$ and the shape operators $\Sh$ and $\Sd$:
\begin{subequations}
\begin{align}
    \label{eq:Gauss parallel}
    \begin{split}
    h[\Phi_\parallel, \Omega^\parallel(X, Y) \Psi_\parallel] &
        = h[\Phi_\parallel, \Omega(X, Y) \Psi_\parallel]
        \\&\qquad
        + h[\Sh(\Phi_\parallel, X), \Sh(\Psi_\parallel, Y)]
        - h[\Sh(\Phi_\parallel, Y), \Sh(\Psi_\parallel, X)]\,,
    \end{split}
    \\
    \label{eq:Gauss normal}
    \begin{split}
    h[\Phi_\perp, \Omega^\perp(X, Y) \Psi_\perp] &
        = h[\Phi_\perp, \Omega(X, Y) \Psi_\perp]
        \\&\qquad
        + h[\Sd(\Phi_\perp, X), \Sd(\Psi_\perp, Y)]
        - h[\Sd(\Phi_\perp, Y), \Sd(\Psi_\perp, X)]\,.
    \end{split}
\end{align}
\end{subequations}
The first equation determines the Berry curvature $\berF = \ii \Omega^\parallel$ in terms of $\Omega$ and the shape operator, while the second equation represents a generalization of the Ricci equation for submanifolds \cite{1999_Spivak_4,2019_Dajczer}. In particular, note that the dependence of the Berry curvature on $\Omega$ is a direct consequence of the non-flat connection $\nabla$ that we assumed, in contrast to other works \cite{PhysRevB.81.245129,mitscherling2024gauge,avdoshkin2024geometry} which start with a flat connection. Working in this more general setting is essential for certain applications, such as semiclassical and high-frequency approximations for wave packets propagating on curved spacetime, where such additional terms arise in the Berry curvature \cite{GSHE2020,Harte_2022,GSHE_GW,PhysRevD.107.044029}.

The Gauss equations can be derived as follows.
Differentiating $\D_Y \Psi_\parallel = P \nabla_Y \Psi_\parallel$ along $X$, one obtains, using \cref{eq:Weingarten},
\begin{align}
    \D_X \D_Y \Psi_\parallel
        = P(\nabla_X \nabla_Y \Psi_\parallel)
        + P (\nabla_X P) \Sh_Y \Psi_\parallel
        + P (\nabla_X P) \D_Y \Psi_\parallel
        \,,
\end{align}
where we have used the notation $\Sh_Y \Psi_\parallel = \Sh(\Psi_\parallel, Y)$ introduced above.
Now, the last term vanishes since $\D_Y \Psi_\parallel$ takes values in $E_\parallel$, causing $(\nabla_X P) \D_Y \Psi_\parallel$ to take values in $E_\perp$, which is precisely the kernel of $P$.
In the second term, $\Sh_Y \Psi_\parallel$ takes values in $E_\perp$, so $P (\nabla_X P)$ can be replaced by $\Sd_X$ there.
This implies
\begin{align}
    h(\Phi_\parallel, \D_X \D_Y \Psi_\parallel)
        = h(\Phi_\parallel, \nabla_X \nabla_Y \Psi_\parallel)
        + h(\Sh_X \Phi_\parallel, \Sh_Y \Psi_\parallel)
        \,.
\end{align}
Anti-symmetrizing over $X$ and $Y$, and using $h(\Phi_\parallel, \D_{[X,Y]} \Psi_\parallel) = h(\Phi_\parallel, \nabla_{[X,Y]} \Psi_\parallel)$, one obtains \cref{eq:Gauss parallel} using the Ricci identity \eqref{eq:Ricci identity}.
The derivation of \cref{eq:Gauss normal} is completely analogous: one replaces $\parallel$ by $\perp$ and $P$ by $P'$; the difference in signs in \cref{eq:Weingarten} plays no role since $\Sd$ enters the final equation quadratically.

In index notation, \cref{eq:Gauss parallel,eq:Gauss normal} take the form
\begin{subequations}
\begin{align}
    \label{eq:Gauss parallel index}
    \t*\Omega{^\parallel_A_B_\mu_\nu}
        &= \t\Omega{_A_B_\mu_\nu} + \t h{_I_J} \{\t{\bar S}{^I_A_\mu} \t S{^J_B_\nu} - \t{\bar S}{^I_A_\nu} \t S{^J_B_\mu} \}\,,
    \\
    \label{eq:Gauss normal index}
    \t*\Omega{^\perp_I_J_\mu_\nu}
        &= \t\Omega{_I_J_\mu_\nu} + \t h{_A_B} \{ \t{\bar S}{^A_I_\mu} \t S{^B_J_\nu} - \t {\bar S}{^A_I_\nu} \t S{^B_J_\mu} \}\,.
\end{align}
\end{subequations}

\subsubsection{Codazzi–Mainardi equations}

As can be seen above, the Gauss equations \eqref{eq:Gauss parallel index} and \eqref{eq:Gauss normal index} allow computing the projections of $\Omega(X,Y)$ onto $E_\parallel$ and $E_\perp$. Similarly, the Codazzi–Mainardi equations express the “mixed components” $\t\Omega{^I_A}(X,Y)$ and $\t\Omega{^A_I}(X,Y)$ in terms of “exterior covariant derivatives” of the shape operators:
\begin{subequations}
\begin{align}
    \label{eq:Codazzi parallel}
    \Dd_X \Sh(\Phi_\parallel, Y) - \Dd_Y \Sh(\Phi_\parallel, X)
        &= + P'[\Omega(X, Y) \Phi_\parallel]\,,
    \\
    \label{eq:Codazzi normal}
    \Dd_X \Sd(\Psi_\perp, Y) - \Dd_Y \Sd(\Psi_\perp, X)
        &= - P[\Omega(X, Y) \Psi_\perp]\,,
\end{align}
\end{subequations}
where $\Dd$ is defined by means of \cref{eq:derivative of shape operator}, in which $\del$ is \emph{any} torsion-free connection on the base manifold $M$. These equations can be derived as follows.

By definition \eqref{def:shape operator 1}, one has $\Sh(\Phi_\parallel, Y) = (\nabla_Y P) \Phi_\parallel = P' \nabla_Y \Phi_\parallel$. Differentiating along $X$ and using \cref{eq:Weingarten}, one obtains
\begin{align}
    \nabla_X [\Sh(\Phi_\parallel, Y)] &
        = P'[\nabla_X \nabla_Y \Phi_\parallel] -  (\nabla_X P) \D_Y \Phi_\parallel - (\nabla_X P) \Sh(\Phi_\parallel, Y)\,.
\end{align}
Applying $P'$ removes the last term, as $\Sh(\Phi_\parallel, Y)$ is a section of $E_\perp$ and is thus mapped by $\nabla_X P$ to a section of $E_\parallel = \kernel P'$. Hence, one finds
\begin{align}
    \D'_X [\Sh(\Phi_\parallel, Y)]
        = P'[\nabla_X \nabla_Y \Phi_\parallel] -  \Sh(\D_Y \Phi_\parallel, X)\,.
\end{align}
From \cref{eq:derivative of shape operator} one then obtains
\begin{align}
    \begin{split}
        [\Dd_X \Sh](\Phi_\parallel, Y) &
        = P'[\nabla_X \nabla_Y \Phi_\parallel]
        - \Sh(\Phi_\parallel, \del_X Y)
        - \Sh(\D_X \Phi_\parallel, Y)
        - \Sh(\D_Y \Phi_\parallel, X)
        \\&
        = P'[ \nabla_X \nabla_Y \Phi_\parallel - \nabla_{\del_X Y} \Phi_\parallel ]
        - [\Sh(\D_X \Phi_\parallel, Y) + \Sh(\D_Y \Phi_\parallel, X)]
        \,.
    \end{split}
\end{align}
Anti-symmetrizing over $X$ and $Y$, the last two terms cancel.
Moreover, since $\del$ is assumed to be torsion-free, one has $\del_X Y - \del_Y X = [X, Y]$, where, as above, $[ \cdot, \cdot]$ denotes the Lie bracket of vector fields. \Cref{eq:Codazzi parallel} is then obtained using the Ricci identity \eqref{eq:Ricci identity}.
The proof of \cref{eq:Codazzi normal} is analogous and is therefore omitted.

In index notation, \cref{eq:Codazzi parallel,eq:Codazzi normal} can be written as
\begin{subequations}
\begin{align}
    \label{eq:Codazzi parallel index}
    \t\Dd{_\mu} \t S{^I_A_\nu} - \t\Dd{_\nu} \t S{^I_A_\mu}
        &= + \t\Omega{^I_A_\mu_\nu}\,,
    \\
    \label{eq:Codazzi normal index}
    \t\Dd{_\mu} \t S{^A_I_\nu} - \t\Dd{_\nu} \t S{^A_I_\mu}
        &= -\t\Omega{^A_I_\mu_\nu}\,.
\end{align}
\end{subequations}

\subsection{Quantum metric}
\label{s:quantum metric}

In this section, we show how a general form of the quantum metric can be obtained starting from the shape operator. After giving a general formula that covers both the Abelian and non-Abelian case, we show how the previous results of Refs.~\cite{Provost1980, PhysRevB.81.245129} can be recovered as particular cases.

For any given $\Phi \in \Gamma(E_\parallel)$ and $X \in \vflds(M)$, the deviation of $D_X \Phi$ from $\nabla_X \Phi$ is measured by the quantity
\begin{align}
    \label{eq:quantum squardnorm}
    q(\Phi, X)
        = h[\t D{_X} \Phi - \t\nabla{_X} \Phi, \t D{_X} \Phi - \t\nabla{_X} \Phi]
        = h[S(\Phi, X), S(\Phi, X)]\,.
\end{align}
Note that the resulting expression depends only algebraically on $\Phi$ and $X$ – the derivatives of $\Phi$ cancel. This means that $q(\Phi, X)$ can be evaluated pointwise without the need to evaluate derivatives of sections.

Evidently, if $\t D{_X} \Phi = \t\nabla{_X} \Phi$ then $q(\Phi, X) = 0$.
Moreover, if $h$ is positive definite, then the converse also holds, that is, $q(\Phi, X) = 0$ implies $\t D{_X} \Phi = \t\nabla{_X} \Phi$.
Regardless of this last property, $q$ defines a quadratic form in the sense that $q(f \Phi, \lambda X) = |f|^2 \lambda^2 q(\Phi, X)$.
This object may be extended to a symmetric tensor as follows.
First, define $\tilde q(\Phi, X, Y)$ to be the unique map that is bilinear in $X$ and $Y$ satisfying $\tilde q(\Phi, X, X) = q(\Phi, X)$ for all $X$ and $\Phi$. Specifically, this is given by the real polarization formula
\begin{align}
    \begin{split}
        \tilde q(\Phi; X, Y) &
            = \tfrac{1}{4}[ q(\Phi, X + Y) - q(\Phi, X - Y) ]
            \\&
            = \tfrac{1}{2} \{ h[\Sh_X \Phi, \Sh_Y \Phi] + h[\Sh_Y \Phi, \Sh_X \Phi] \}\,.
    \end{split}
\end{align}
Finally, $\tilde q(\Phi, X, Y)$ extends uniquely to a tensorial expression $G(\Phi, \Psi; X, Y)$ that is sesquilinear in $\Phi$ and $\Psi \in \Gamma(E_\parallel)$, bilinear in $X$ and $Y$, and satisfies $G(\Phi, \Phi; X, Y) = \tilde q(\Phi, X, Y)$. Indeed, this quantity is given by the complex polarization formula:
\begin{align}
    \begin{split}
        G(\Phi, \Psi; X, Y) &
            = \tfrac{1}{4} [ + \tilde q(\Phi + \Psi; X, Y) - \tilde q(\Phi - \Psi; X, Y)
            - \ii \tilde q(\Phi + \ii \Psi; X, Y) + \tilde q(\Phi - \ii \Psi; X, Y)]
            \\&
            = \tfrac{1}{2} \{ h[\Sh_X \Phi, \Sh_Y \Psi] + h[\Sh_Y \Phi, \Sh_X \Psi] \}\,.
    \end{split}
\end{align}
The tensor $G$ is known as the quantum metric. In index notation, it takes the form
\begin{align}
    \label{eq:quantum metric indices}
    \t G{_A_B_\mu_\nu}
        = \half \t h{_I_J}\{
            \t{\bar S}{^I_A_\mu} \t S{^J_B_\nu}
            + \t{\bar S}{^I_A_\nu} \t S{^J_B_\mu}
        \}\,.
\end{align}
Evidently, $G$ has the symmetry properties
\begin{align}
    \t G{_A_B_\nu_\mu} &= \t G{_A_B_\mu_\nu}
    &
    \t G{_B_A_\mu_\nu} &= \t{\bar G}{_A_B_\mu_\nu}\,,
\end{align}
so $G$ may be regarded as a symmetric Hermitian-matrix-valued covariant $2$-tensor.

The quantum metric $\t G{_A_B_\mu_\nu}$ induces a symmetric $2$-tensor $\t G{_\mu_\nu} = \t h{^A^B} \t G{_A_B_\mu_\nu}$ (where $\t h{^A^B}$  is the inverse of $\t h{_A_B}$) on the base manifold $M$. If $\t G{_\mu_\nu}$ is non-degenerate, it defines a Riemannian or pseudo-Riemannian metric on $\mfd$ (with its associated Levi-Civita derivative).
However, in the following, no such non-degeneracy assumption will be made.
As a simple example, and also as a consistency check, it is straightforward to show that our general result can reproduce the Abelian quantum metric introduced in Refs.~\cite{Provost1980,PhysRevB.81.245129}. In this case, a flat connection $\nabla = \partial$ is assumed on $E$, and the sub-bundle geometry is determined by a projector $P$ of rank $1$. In particular, using the Dirac notation, the projector can be locally (or even globally if we further assume that $E = M \times \numC^n$) represented as $P = \ketbra{\Psi}{\Psi}$, where $\braket{\Psi \mid \Psi} = 1$. For example, given a local frame $(s_i)$, the wave function $\Psi$ can be identified with a particular element of the frame. With these assumptions, the shape operator can be expressed as
\begin{equation}
    \Sh(\Psi, X)
        = (\partial_X P) \Psi
        = \t X{^\mu} \left( \ket{\partial_\mu \Psi} + \braket{ \partial_\mu \Psi \mid \Psi} \ket{\Psi} \right)\,.
\end{equation}
Using this result, the quantum metric becomes $G(\Psi, \Psi; X, Y) = X^\mu Y^\nu \t G{_{\mu \nu}}$, where
\begin{equation}
    \t G{_\mu_\nu}
        = \re \braket{\partial_\mu \Psi \mid \partial_\nu \Psi} + \braket{ \Psi \mid \partial_\mu \Psi} \braket{ \Psi \mid \partial_\nu \Psi}
        = \re \braket{ \partial_\mu \Psi \mid I - P \mid \partial_\nu \Psi }\,.
\end{equation}
This result coincides with the definition of the quantum metric given in Ref.~\cite[Eq.~(2.12)]{Provost1980}, as well as the one given in Ref.~\cite[Eq.~(6)]{PhysRevB.81.245129}. In this case, the quantum metric tensor $\t G{_\mu_\nu}$ defines a way of measuring the distance between quantum states along paths in the parameter space. However, note that $\t G{_\mu_\nu}$ is only positive semi-definite, and therefore not a metric on parameter space \cite[Sec.~2.2.6]{Geometric_phases_book}. On the other hand, the quantum metric $\t G{_\mu_\nu}$ corresponds to the Fubini–Study metric on the projective Hilbert space $\numC P^{n-1}$ \cite[Sec.~5.3.2]{Geometric_phases_book}.

\subsection{Quantum geometric tensor}

Based on the equations obtained above for the Berry curvature and the quantum metric in terms of the shape operator, we can define the quantum geometric tensor $Q$ as a direct tensorial extension of the quantity $q$ introduced in \cref{eq:quantum squardnorm}, namely
\begin{align}
    \label{eq:QGT definition}
    Q(\Phi, \Psi; X, Y)
        = h[S(\Phi, X), S(\Psi, Y)]
        \equiv h[(\nabla_X P) \Phi, (\nabla_Y P) \Psi]\,,
\end{align}
where $\Phi, \Psi \in E_\parallel = \image P$.
In index notation, one thus has
\begin{align}
    \t Q{_A_B_\mu_\nu}
        &= \t h{_I_J} \t{\bar S}{^I_A_\mu} \t S{^J_B_\nu}\, ,
\end{align}
and $Q$ can be seen as a direct generalization of the third fundamental form of classical differential geometry \cite{2019_Dajczer}. A similar form of the quantum geometric tensor, although in a simplified geometric setting, was also given in Refs.~\cite{PhysRevB.89.235424,Northe,Romero_2024}, where the shape operators are referred to as “$N$-beins” or “zweibeins” which behave as a “square root” of the quantum geometric tensor, similarly to how the spacetime metric in general relativity can be expressed in terms of tetrad fields.

The quantum geometric tensor satisfies
\begin{align}
    \t Q{_A_B_\mu_\nu}
        &= \t{\bar Q}{_B_A_\nu_\mu}\,,
\end{align}
so that symmetrization and anti-symmetrization over the last two indices coincide with taking the Hermitian or anti-Hermitian part in the first two indices, respectively.
Specifically, decomposing $\t Q{_A_B_\mu_\nu}$ as
\begin{align}
    \t Q{_A_B_\mu_\nu}
        = \t Q{_A_B_(_\mu_\nu_)} + \t Q{_A_B_[_\mu_\nu_]}\,,
\end{align}
where parentheses indicate symmetrization and brackets denote anti-symmetrization, \cref{eq:quantum metric indices,eq:Gauss parallel index} show that
\begin{subequations}
\begin{align}
    \t Q{_A_B_(_\mu_\nu_)}
        &= \t G{_A_B_\mu_\nu}\,,
    \\
    \t Q{_A_B_[_\mu_\nu_]}
        &= \half( \t*\Omega{^\parallel_A_B_\mu_\nu} - \t*\Omega{_A_B_\mu_\nu} )\,.
\end{align}
\end{subequations}
Therefore, we can write the quantum geometric tensor as
\begin{align}
    \label{eq:QGT decomposition}
    \begin{split}
        \t Q{_A_B_\mu_\nu} &
            = \t G{_A_B_\mu_\nu} + \half( \t*\Omega{^\parallel_A_B_\mu_\nu} - \t*\Omega{_A_B_\mu_\nu} ) = \t G{_A_B_\mu_\nu} - \ihalf \t*\berF{_A_B_\mu_\nu} - \half \t*\Omega{_A_B_\mu_\nu}\,.
    \end{split}
\end{align}
This generalizes previous results (such as Ref.~\cite{PhysRevB.81.245129}) by including the last term that accounts for the curvature $\Omega$ of the connection $\nabla$. Regarding the physical role of this additional curvature contribution, it is well known that the Berry curvature $\berF$ determines the semiclassical dynamics of wave packets in condensed matter physics \cite{Niu,Niu2}. In this case, the connection is flat and there is no additional curvature contribution to $Q$. However, for wave packets propagating in curved spacetime, the gravitational spin Hall effects \cite{GSHE2020,GSHE_GW,Harte_2022,PhysRevD.107.044029} are determined by the antisymmetric part of $Q$, containing both the Berry curvature $\berF$ and the additional curvature contribution $\Omega$. This leads to a spin–gravity coupling and anomalous transport phenomena.

Using the properties of the projector $P$, the quantum geometric tensor can also be written in various equivalent forms, such as
\begin{align}
    \label{eq:Q tensor alternative formulas}
    \begin{split}
    Q(\Phi, \Psi; X, Y)&
        = h[\nabla_X \Phi, (I-P) \nabla_Y \Psi] = h[\Phi, (\nabla_X P) (\nabla_Y P) \Psi]\,.
    \end{split}
\end{align}
The first expression here corresponds to Eq.~(9) of Ref.~\cite{PhysRevB.81.245129}, and the second expression is used in Refs. \cite{PhysRevB.107.245136,avdoshkin2024geometry,mitscherling2024gauge}.

\section{Applications}
\label{s:applications}

In this section, we apply the geometric concepts introduced above to the semiclassical approximation of the Dirac equation on a general spacetime \cite{PhysRevD.107.044029}. Working with a non-flat connection is essential in this case.
We show how the semiclassical expansion defines a projector $P$. Then, all the geometric objects describing the sub-bundle geometry can be defined as in the previous section. The quantum geometric tensor, together with the quantum metric and the Berry curvature, are then tensors on phase space. In particular, the Berry curvature contains a Riemann tensor contribution and can be viewed as a generalization of the static phase-space Berry curvature discussed in Refs.~\cite{Niu,Niu2,Hayata}.
Finally, using these semiclassical results, we examine the particular case of Dirac fermions confined on the hyperbolic plane \cite{Comtet,Gorban,Thiang}. This can be seen as a continuum limit for hyperbolic Dirac materials \cite{Tummuru,Ikeda,Roy2,Roy3}. We show how the geometry of the hyperbolic plane has a direct impact on the quantum metric.

\subsection{Semiclassical dynamics of Dirac fields in curved spacetime} \label{sec:Dirac}

Our main goal here is to show how the quantum geometric tensor, together with the Berry curvature and quantum metric, can be defined for semiclassical Dirac fields in curved spacetime. The semiclassical regime can be described using a WKB approximation \cite{PhysRevD.107.044029}, and the sub-bundle geometry arises from the eigenspaces of the principal symbol of the Dirac operator (this is generally the case for WKB-type approximations of multicomponent fields \cite{Emmrich1996}). Note that, compared to \cref{s:geometric setup,s:projected connections}, some additional geometric structure is implicitly assumed here. In particular, we assume a smooth Lorentzian manifold $(M, g)$, as well as the Levi-Civita connection $\Gamma(g)$. Our choice of signature for the metric $g$ is $(- + + +)$. Furthermore, the existence of a spin structure is also assumed, so that spinor fields are well-defined on $M$.

The Dirac fields $\Psi$ of charge $q$ and mass $m$ are sections of a complex vector bundle with typical fiber $\numC^4$ that satisfy the Dirac equation
\begin{equation} \label{eq:Dirac_eq}
    \left(
        \ii \hbar \t\gamma{^\mu} \t{\nabla\!}{_\mu}
        + q \t\gamma{^\mu} \t A{_\mu}
        - m
    \right) \Psi = 0\,,
\end{equation}
where $\t{\nabla\!}{_\mu}$ is the spinor covariant derivative (defined by the pullback of the Levi-Civita connection with the spin structure \cite[p. 419]{MR685274}), $\t A{_\mu}$ is the electromagnetic potential and $\t\gamma{^\mu}$ are the spacetime gamma matrices. These are related to the flat spacetime gamma matrices $\t\gamma{^a}$ through a choice of tetrad field $(e_a)^\mu$ as $\gamma^\mu = (e_a)^\mu \gamma^a$, which implies $\t\gamma{^\mu} \t\gamma{^\nu} + \t\gamma{^\nu} \t\gamma{^\mu} = -2 \t g{^\mu^\nu}$. The action of the spinor covariant derivative on a Dirac field is
\begin{equation}
    \t{\nabla\!}{_\mu} \Psi
        = \t\p{_\mu} \Psi
        - \tfrac{1}{4} \t\omega{^a^b_\mu} \t\gamma{_a} \t\gamma{_b} \Psi\,,
\end{equation}
where the spin connection coefficients are defined as $\t\omega{^a^b_\mu} = (\t e{^a}){_\nu} \t{\nabla\!}{_\mu} (\t e{^b}){^\nu}$.
The corresponding curvature is thus related to the Riemann tensor $\t R{_\mu_\nu_\rho_\sigma}$ via
\begin{align}
    \Omega(X, Y)
        = - \tfrac{1}{4} \t R{_\alpha_\beta_\mu_\nu} \t X{^\mu} \t Y{^\nu} \t\gamma{^\alpha} \t\gamma{^\beta}
        = \tfrac{\ii}{4} \t R{_\alpha_\beta_\mu_\nu}\t X{^\mu} \t Y{^\nu} \t \sigma{^\alpha^\beta}\,,
\end{align}
where $\t\sigma{^\alpha^\beta} = \tfrac{\ii}{2}[\t\gamma{^\alpha}, \t\gamma{^\beta}]$.
The semiclassical expansion of the Dirac equations is obtained from a WKB ansatz of the form
\begin{subequations} \label{eq:WKB_Dirac}
\begin{align}
    \Psi (x)
        &= \psi(x, \nabla_\mu S, \hbar) \, \ee^{ \ii S(x) / \hbar}\,,
    \\
    \psi(x,  \nabla_\mu S, \hbar)
        &= {\psi_0}(x,  \nabla_\mu S)
        + \hbar {\psi_1}(x,  \nabla_\mu S)
        + \mathcal{O}(\hbar^2)\,,
\end{align}
\end{subequations}
where $S$ is a real scalar function, $\psi$ is a complex amplitude spinor, and Planck's constant $\hbar$ is regarded as a small expansion parameter. The semiclassical equations are obtained by inserting the WKB ansatz into the Dirac equation and setting to zero the resulting terms at each order in $\hbar$.

\subsubsection{Leading-order equations}

At leading order in the WKB expansion, we obtain a homogeneous system of linear algebraic equations
\begin{equation}
    \Dir \psi_0 = 0\,,
\end{equation}
where $\Dir = - \t k{_\mu} \t\gamma{^\mu} - m$ is the principal symbol of the Dirac operator, and $\t k{_\mu} = \t{\nabla\!}{_\mu} S - q \t A{_\mu}$. This system of equations admits non-trivial solutions only if the determinant of the principal symbol vanishes, which defines the dispersion relation:
\begin{align} \label{eq:disp}
    \det( \Dir ) &= 0
    &
    &\Leftrightarrow
    &
    \t g{^\mu^\nu} \t k{_\mu} \t k{_\nu} &= - m^2\,.
\end{align}
The dispersion relation is a Hamilton–Jacobi equation for the phase function $S$, which can be solved using the method of characteristics \cite[Sec.~46]{1989_Arnold}. By differentiating the dispersion relation, it also follows that $\t k{_\mu}$ satisfies the Lorentz force law
\begin{equation} \label{eq:Lorentz_force}
    \t k{^\nu} \t\nabla{_\nu} \t k{_\mu}
        = q \t F{_\mu_\nu} \t k{^\nu} \,.
\end{equation}

With the dispersion relation imposed, $\Dir$ has the spectral decomposition $\Dir = 0 P - 2 m P'$, where $P$ and $P'$ are the projectors onto the kernel of $\Dir$ and its orthogonal complement:
\begin{align}
    \label{eq:Dirac projectors explicit}
    P &= \tfrac{1}{2m}(m - \t k{_\mu} \t\gamma{^\mu})\,,
    &
    P' &= \tfrac{1}{2m}(m + \t k{_\mu} \t\gamma{^\mu})\,.
\end{align}
Furthermore, from the WKB expansion of the conserved Dirac current $j^\mu = \braket{\Psi \mid \gamma^\mu \Psi}$ and $\nabla_\mu j^\mu = 0$, we obtain a transport equation for the zeroth-order intensity $\mathcal{I} = \braket{\psi_0 \mid \psi_0}$ of the WKB field:
\begin{equation} \label{eq:transp_I}
    \t{\nabla\!}{_\mu} \t j{_0}^\mu
        = \t{\nabla\!}{_\mu} \braket{\psi_0 \mid \t\gamma{^\mu} \psi_0}
        = \tfrac{1}{m} \t{\nabla\!}{_\mu} \left( \mathcal{I} \t k{^\mu} \right) = 0\,.
\end{equation}
To summarize, from the lowest order in the WKB expansion we obtained the dispersion relation, a transport equation for the zeroth order intensity of the field, and the eigenspace in which the amplitude $\psi_0$ is supposed to be valued. Thus, the most general form that $\psi_0$ can take is
\begin{equation}
    \psi_0 = \sqrt{\mathcal{I}}\, \psi,
\end{equation}
with $P \psi = \psi$ and $ \braket{\psi \mid \psi} = 1$.
To obtain further information about $\psi$, it is necessary to extend the WKB expansion to the next order.

\subsubsection{Transport equation and Berry connection}

At next-to-leading order in the WKB expansion, the Dirac equation implies
\begin{equation}
    \Dir \psi_1  = - \ii  \gamma^\mu \nabla_\mu \psi_0\,.
\end{equation}
Since $\Dir$ is not invertible, the condition that the inhomogeneity on the right-hand side is in the image of $\Dir$ constitutes a non-trivial constraint, namely
\begin{equation}
    P \t\gamma{^\mu} \t{\nabla\!}{_\mu} \psi_0 = 0\,.
\end{equation}
Using $\psi_0 = P \psi_0$, this can be written in the form
\begin{align}
    (P \t\gamma{^\mu} P) \t{\nabla\!}{_\mu} \psi_0
        = - P \t\gamma{^\mu} (\t{\nabla\!}{_\mu} P) \psi_0\,.
\end{align}
Furthermore, using the relation $P \t\gamma{^\mu} P = P \t k{^\mu} / m $, we obtain
\begin{align}
    \t k{^\mu} \t D{_\mu} \psi_0
        = - m P \t\gamma{^\mu} (\t\nabla{_\mu} P) \psi_0\,,
\end{align}
where $\t{D\!}{_\mu} = P \t{\nabla\!}{_\mu}$ is the projected connection.
Using the explicit form of $P$ given in \cref{eq:Dirac projectors explicit}, as well as the identity $\t\gamma{^\mu} \t\gamma{^\nu} = - \t g{^\mu^\nu} - \ii \t\sigma{^\mu^\nu}$, one finds
\begin{align}
    m \t\gamma{^\mu} \t{\nabla\!}{_\mu} P &
        = - \half \t\gamma{^\mu} \t\gamma{^\nu} \t{\nabla\!}{_\mu} \t k{_\nu}
        = \half \t g{^\mu^\nu} \t{\nabla\!}{_\mu} \t k{_\nu}
        + \ihalf \t\sigma{^\mu^\nu} \t{\nabla\!}{_[_\mu} \t k{_\nu_]}
        = \half \t{\nabla\!}{_\mu} \t k{^\mu}
        - \tfrac{\ii}{4} q \t\sigma{^\mu^\nu} \t F{_\mu_\nu}\,,
\end{align}
where the last step follows from $\t k{_\mu} = \t{\nabla\!}{_\mu} S - q \t A{_\mu}$.
As a consequence, the transport equation for $\psi_0$ takes the form
\begin{align} \label{eq:transport_eq}
    \t k{^\mu} \t D{_\mu} \psi_0
        = - \half \left(\t \nabla{_\mu} \t k{^\mu} \right) \psi_0
        + \tfrac{\ii}{4} q \t F{_\mu_\nu} P \t\sigma{^\mu^\nu} \psi_0
        \,.
\end{align}
Factoring $\psi_0 = \sqrt{\mathcal I}\, \psi$ and using the transport equation \eqref{eq:transp_I} for the intensity, one obtains the transport equation for the normalized spinor $\psi$:
\begin{equation} \label{eq:transp_berry}
    k^\mu D_\mu \psi
        = \tfrac{\ii}{4} q F_{\mu \nu} P \sigma^{\mu \nu} \psi\,.
\end{equation}
The term on the right-hand side is the so-called “no-name” term introduced in Ref.~\cite{PhysRevA.44.5239} (see also Ref.~\cite{PhysRevD.91.025004}). Its geometric origin has been discussed in Ref.~\cite{Emmrich1996}.

The projected connection is directly related to the Berry connection by a factor of $\ii$. This can be seen by writing the connection form with respect to the frame $\t s{_A}$:
\begin{align}
    \t k{^\mu} \t{\D\!}{_\mu} \t s{_A}
        = \t k{^\mu} P \t{\nabla\!}{_\mu} \t s{_A}
        = \t k{^\mu} \t \Gamma{^B_A_\mu} \t s{_B}\,,
\end{align}
where $\t \Gamma{^B_A_\mu} = \t\sigma{^B}(\t{\nabla\!}{_\mu}\t s{_A})$ and hence $\t \Gamma{_B_A_\mu} = \braket{\t s{_B} \mid \t{\nabla\!}{_\mu} \t s{_A}}$.
The usual expression of the Berry connection form is then obtained as $\t{\mathcal{B}}{_B_A_\mu} = \ii \t\Gamma{_B_A_\mu} = \ii \braket{\t s{_B} \mid \t{\nabla\!}{_\mu} \t s{_A}}$, which is the same as in Ref.~\cite[Eq. (4.35a)]{PhysRevD.107.044029}. The imaginary unit makes the Berry connection a Hermitian matrix-valued one-form, with the matrix part corresponding to the $\mathfrak{u}(2)$ Lie algebra.

Note also that the transport equation does not hold along arbitrary vector fields on spacetime, but just along vector fields $\t k{^\mu}$ satisfying the dispersion relation \eqref{eq:disp} and the Lorentz force law \eqref{eq:Lorentz_force}. In other words, solutions of the Hamilton-Jacobi equation \eqref{eq:disp} are obtained through the method of characteristics by integrating the phase function $S$ along the rays determined by the Hamiltonian
\begin{align}
    H(x, p)
        = \half g^{\mu \nu} ( p_\mu - q A_\mu ) ( p_\nu - q A_\nu )\,.
\end{align}
Then, the corresponding Hamiltonian vector field with respect to the canonical symplectic form $\dd x^\mu \wedge \dd p_\mu$ is
\begin{align}
    \begin{split}
        X_H &
            = \t{\dot x}{^\mu} \frac{\p}{\p \t x{^\mu}}
                + \t{\dot p}{_\mu} \frac{\p}{\p \t p{_\mu}}
            \\&
            = (\t p{^\mu} - q \t A{^\mu}) \frac{\p}{\p \t x{^\mu}}
            + [\t\Gamma{^\rho_\nu_\mu} (\t p{_\rho} - q \t A{_\rho}) + q \t\p{_\mu} \t A{_\nu}] (\t p{^\nu} - q \t A{^\nu})  \frac{\p}{\p \t p{_\mu}}
            \,.
    \end{split}
\end{align}
The spacetime vector field $k^\mu$ is obtained by first projecting $X_H$ to the Lagrangian submanifold $L \subset T^*M$ [where $L$ is defined as $H^{-1}(\{-m^2/2\}) |_{p_\mu = \nabla_\mu S}$]:
\begin{align}
    X_L
        = X_H \big|_L
        = \t k{^\mu} \frac{\p}{\p \t x{^\mu}}
            + ( \t\Gamma{^\rho_\nu_\mu} \t k{_\rho} \t k{^\nu} + q \t k{^\nu} \t\p{_\mu} \t A{_\nu}) \frac{\p}{\p \t p{_\mu}} \,,
\end{align}
and then projecting from $L$ to $M$.

The spinor $\psi$ is also naturally defined on $L$ since $\Dir = \Dir(x^\mu, \nabla_\mu S)$ and $\Dir \psi = 0$. Thus, one generally has eigenspinors with the functional dependence $\psi = \psi(x^\mu, \nabla_\mu S)$. This justifies taking the Lagrangian submanifold $L$ with local coordinates $(x^\mu, \nabla_\mu S)$ as the parameter space for the transport equation \eqref{eq:transport_eq}, as is generally the case for transport equations resulting from a WKB expansion \cite{Emmrich1996}.

\Cref{eq:transp_berry} can be reformulated as a transport equation on the Lagrangian submanifold $L$ as discussed in Ref.~\cite[Sec.~IV.D]{PhysRevD.107.044029}. First, since $\psi = \psi\left[x^\mu, \nabla_\mu S(x) \right]$, we need to apply the chain rule when taking derivatives in the transport equation:
\begin{align}
    \begin{split}
        \t k{^\mu} \t{\D\!}{_\mu} [\psi(x, \nabla S)] &
            = \t k{^\mu} (\t{\D\!}{_\mu} \psi)(x, \nabla S)
            + \t k{^\mu} (\t\p{_\mu} \t{\nabla\!}{_\nu} S)
                \left( P \frac{\p \psi}{\p \t{\nabla\!}{_\nu} S} \right)(x, \nabla S)
            \\&
            = \t k{^\mu} (\t{\D\!}{_\mu} \psi)(x, \nabla S)
            + \t k{^\mu} (\t\Gamma{^\rho_\mu_\nu} \t k{_\rho} + q \t A{_\mu}  \t p{_\nu})
                \left( P \frac{\p \psi}{\p \t{\nabla\!}{_\nu} S} \right)(x, \nabla S)
            \,.
    \end{split}
\end{align}
In the above equation, the components of the Hamiltonian vector field $X_L$ are contracted into the two derivatives, so one can write
\begin{align}
    \t k{^\mu} \t{\D\!}{_\mu} [\psi(x, \nabla S)] = \tilde{D}_{X_L} \psi\,,
\end{align}
where the connection on the Lagrangian submanifold is defined as $\tilde{D} = D + P \frac{\partial}{\partial \nabla S}$. The corresponding Berry connection is now a Lie algebra-valued one-form on $L$ and takes the form
\begin{subequations} \label{eq:transp_L}
\begin{align}
    \ii \tilde{D}_{X_L} s_A
        &= \t{{\mathcal B}}{^B_A}(X_L) s_B\,,
    \\
    \t{{\mathcal B}}{_B_A}
        &= \ii \braket{ s_B \mid \t{\nabla\!}{_\mu} s_A} \t{\dd x}{^\mu}
        + \ii \braket{ s_B \mid  \frac{\p}{\p \t{\nabla\!}{_\mu} S} s_A} \t{\dd p}{_\mu}\,.
\end{align}
\end{subequations}

\subsubsection{Quantum geometric tensor}

The quantum geometric tensor, as well as the Berry curvature and the quantum metric, are objects that act on two fields and two vectors in the parameter space. In the WKB construction presented above, the amplitudes are naturally defined on a Lagrangian submanifold of $T^*M$, but in general the amplitudes of different WKB fields are defined on different Lagrangian submanifolds. For example, two WKB states propagating in different directions necessarily have different phase functions $S$.
Thus, for the quantum geometric tensor to be able to act on all possible WKB states, it must be defined on a larger bundle that contains all Lagrangian submanifolds arising the WKB construction, namely
\begin{align}
    \varGamma = \{
        p \in T^*M \mid
        \t g{^\mu^\nu} \t p{_\mu} \t p{_\nu} = -m^2
        \text{ and }
        \t p{^\mu} \text{ is future-directed}
    \}\,.
\end{align}
Denoting by $\spinorsB \overset{\spinorsP}{\to} M$ the spinor bundle over spacetime (with typical fiber $\numC^4$) and by $\pi$ the canonical projection of $\varGamma$ to $M$, the quantum geometric tensor can be constructed naturally over the pullback bundle $E = \pi^* \spinorsB$.
Via the pullback, the fiber metric $h$ and the compatible covariant derivative $\nabla$ on $\spinorsB$ induce similar structure on $E$ — in the following, these quantities will also be denoted by $h$ and $\nabla$ for notational simplicity.

The key aspects of the above semiclassical analysis carry over to $\varGamma$ as most of the equations hold for general momenta $\t p{_\mu}$ and not only those of the form $\t k{_\mu} = \t{\nabla\!}{_\mu} S - q \t A{_\mu}$ (exceptions are \cref{eq:Lorentz_force,eq:transport_eq} which require $\t{\nabla\!}{_\mu}\t k{_\nu} - \t{\nabla\!}{_\nu}\t k{_\mu} = - q \t F{_\mu_\nu}$).
Notably, our construction shows that the quantum geometric tensor is fully determined by the bundle $E$ (with its structures $h$ and $\nabla$) and the projector $P$, and these geometric structures are unambiguously defined on $E = \pi^* \spinorsB$. In particular, the eigenvalue problem $\Dir \psi = 0$ is well-defined on $E$ with $\Dir = \Dir(x, p)$.
As above, $\Dir$ has the spectral decomposition $\Dir = 0 P - 2 m P'$ with $P$ and $P'$ given by
\begin{align}
    P &= \tfrac{1}{2m}(m - \t p{_\mu} \t\gamma{^\mu} )\,,
    &
    P' &= \tfrac{1}{2m}(m + \t p{_\mu} \t\gamma{^\mu} )\,,
\end{align}
in complete analogy with \cref{eq:Dirac projectors explicit}.

In this setting, it is trivial to compute the shape operator according to \cref{def:shape operator 1} and the quantum geometric tensor using \cref{eq:QGT definition}.
Explicitly, using the fiber coordinates $(\t x{^\mu}, \t p{_\nu})$, let
\begin{align}
    \t\Pi{_\alpha}
        = \nabla \t p{_\alpha}
        = \t {\dd p}{_\alpha} - \t p{_\rho} \t \Gamma{^\rho_\alpha_\mu} \t{\dd x}{^\mu}\,,
\end{align}
which is a natural quantity for expressing the condition that a vector field $X = \t v{^\mu} \frac{\p}{\p x{^\mu}} + \t w{_\nu} \frac{\p}{\p \t p{_\nu}}$ on $T^*M$ is tangent to $\varGamma$, namely
\begin{align}
    \label{eq:on-shell constraint}
    \t p{^\alpha} \t \Pi{_\alpha}(X) = 0\,.
\end{align}
In terms of $\t\Pi{_\alpha}$, the shape operator takes the form
\begin{align}
    \Sh(\psi, X)
        \equiv \Sh_X \psi
        = - \tfrac{1}{2m} \t \Pi{_\alpha}(X) \, \t\gamma{^\alpha} \psi\,.
\end{align}
\Cref{eq:QGT definition} then shows that the components of the quantum geometric tensor (relative to a frame $\t s{_A}$) are
\begin{align}
    \t Q{_A_B}
        = \tfrac{1}{4 m^2} \braket{\t\gamma{^\alpha} \t s{_A} \mid \t\gamma{^\beta} \t s{_B}} \t\Pi{_\alpha} \otimes \t\Pi{_\beta}\,.
\end{align}
Denoting the symmetric tensor product by $\odot$ and the exterior product by $\wedge$, the quantum metric and the Berry curvature are found to be given by
\begin{subequations}
\begin{align}
    \label{eq:Dirac:G metric}
    \t G{_A_B}
        &= - \tfrac{1}{4} m^{-2}\, \t h{_A_B} \t g{^\alpha^\beta} \t\Pi{_\alpha} \odot \t\Pi{_\beta}\,,
    \\
    \label{eq:Dirac:F curvature}
    \t \berF{_A_B}
        &= + \tfrac{1}{4} \t \varsigma{_A_B^\alpha^\beta} [
            m^{-2}\, \t\Pi{_\alpha} \wedge \t\Pi{_\beta}
            - \half \t R{_\alpha_\beta_\mu_\nu} \t{\dd x}{^\mu} \wedge \t{\dd x}{^\nu}
        ]\,,
\end{align}
\end{subequations}
where $\t h{_A_B} = \braket{\t s{_A} \mid \t s{_B}}$ and $\t \varsigma{_A_B^\alpha^\beta} = \braket{\t s{_A} \mid \t \sigma{^\alpha^\beta} \t s{_B}} = -\tfrac{\ii}{2} \braket{\t s{_A} \mid [\t\gamma{^\alpha}, \t\gamma{^\beta}] \t s{_b}}$.

The same form of the Berry curvature, although written as a tensor on $T^*M$ instead of $\varGamma$, was also given in Ref.~\cite{PhysRevD.107.044029} (and also in Ref.~\cite{PhysRevD.91.025004} for the particular case of Minkowski spacetime). However, the expressions for the quantum geometric tensor and the quantum metric are new.

\subsection{Hyperbolic Dirac materials}

In this section, as an application of our formalism oriented towards condensed matter physics, we will consider massive Dirac fermions on the hyperbolic plane \cite{Comtet,Gorban,Thiang}. This is relevant for understanding the physical behavior of electronic systems on hyperbolic geometries in the low-energy limit, particularly in the context of hyperbolic Dirac materials \cite{Tummuru,Ikeda,Roy2,Roy3}, where hyperbolic lattices govern their electronic structure and have been experimentally realized in synthetic matter \cite{Kollar,Chen3,Huang4}. These systems, among other things, can support topological phases characterized by the Berry connection and curvature effects, offering insights into exotic topological phenomena \cite{Yu,Zhang2,Bzdusek,Bzdusek2,Maciejko2,Bzdusek3}.
Hyperbolic lattices are constructed by tessellating negatively curved spaces with regular polygons. Unlike Euclidean lattices, where the number of sites grows polynomially with distance, hyperbolic lattices exhibit exponential growth in the number of lattice sites as one moves outward from a central site. This unique structure significantly modifies the quantum dynamics of particles on these lattices.
In bipartite hyperbolic lattices (analogous to graphene's honeycomb lattice \cite{Novoselov,Pachos3,Marzuoli2,Pachos2}), massless Dirac fermions emerge near half-filling because of the lattice symmetry and geometry, exhibiting linear dispersion around Dirac points. However, the corresponding density of states is altered compared to flat-space systems, acquiring a finite value near zero energy in the presence of strong magnetic fields \cite{Roy2}.
Furthermore, the introduction of a Dirac mass (gap) modifies the topology of the band structure, potentially leading to new topological phases such as hyperbolic Chern insulators \cite{Zhang2, Maciejko2}. In fact, the interplay between mass and curvature can result in a non-trivial Berry connection that gives rise to an anomalous Hall conductivity \cite{Bzdusek3}.

As we will show below and differently from the previously mentioned works, here we are interested in studying the phase-space quantum metric induced by the non-trivial geometry of the hyperbolic plane.
In both the Poincaré disk model and the Poincaré half-plane model, the metric of the hyperbolic plane takes the form
\begin{align}
    \tilde g
        &= \lambda(x,y)^2 (\dd x^2 + \dd y^2)\,,
\end{align}
with the conformal factor $\lambda$ being $\lambda = 2a/(1 - x^2 - y^2)$ for the disk model (where $x^2 + y^2 < 1$), or $\lambda = a/x$ for the half-space model (where $x > 0$).
Here, $a$ is a constant factor that is related to the scalar curvature via
\begin{align}
    \operatorname{Scal}(\tilde g)
        &= - \frac{2}{a^2}\,.
\end{align}
We now consider the continuum limit of hyperbolic Dirac systems \cite{Tummuru,Ikeda,Roy2,Roy3} and, more specifically, focus on massive fermions in which the Dirac field is given by a four-component spinor \cite{Tummuru}. In the low-energy regime, these systems are described by a (2+1)-dimensional Dirac theory on the hyperbolic plane \cite{Comtet,Gorban,Thiang}.
Thus, we can apply the language of \cref{sec:Dirac} as follows. First, we consider the Dirac equation in the background spacetime with metric
\begin{equation}
    g = -\dd t^2 + \lambda^2(x,y) (\dd x^2 + \dd y^2) + \dd z^2,
\end{equation}
and an external electromagnetic field
\begin{equation} \label{eq:Em}
    A = A_t(t,x,y) \dd t + A_x(t,x,y) \dd x + A_y(t,x,y) \dd y.
\end{equation}
In this case, the Dirac operator is
\begin{align}
    \begin{split}
        \mathcal D &
            = \ii \hbar \t\gamma{^\mu} \t{\nabla\!}{_\mu}
                + q \t\gamma{^\mu} \t A{_\mu}
                - m
            \\&
            = \ii \hbar \t\gamma{^0} \t\p{_t}
            + \ii \hbar \t\gamma{^3} \t\p{_z}
            + q \t\gamma{^\mu} \t A{_\mu}
            - m
            + \frac{\ii \hbar}{\lambda} \t\gamma{^1}(\t\p{_x} + \half \t\p{_x} \ln \lambda)
            + \frac{\ii \hbar}{\lambda} \t\gamma{^2}(\t\p{_y} + \half \t\p{_y} \ln \lambda)
            \,,
    \end{split}
\end{align}
where the last three terms above correspond to the $2$-dimensional Dirac operator on the hyperbolic plane considered in Ref.~\cite{Comtet} (note that this paper uses opposite sign conventions for the anticommutator of gamma matrices and the mass $m$).

To describe Dirac fermions localized on the hyperbolic plane, we consider the following particular form of the WKB ansatz:
\begin{equation}
    \Psi
        = \sqrt{\mathcal{I} (t,x,y)\, \ee^{-z^2/\epsilon}}\,
        \psi(t,x,y) \ee^{\frac{\ii}{\hbar} S(t,x,y) }\,,
\end{equation}
where $\epsilon$ is a constant. This choice of ansatz ensures that a WKB wave packet remains localized on the $xy$-plane, and the external electromagnetic field can be used to control the position of the wave packet in the plane. To see that this is indeed the case, we impose the WKB equations derived in \cref{sec:Dirac}.

The wave vector
\begin{align}
    k_\mu = \begin{pmatrix} k_t \\ k_x \\ k_y \\ k_z
    \end{pmatrix} =  \begin{pmatrix} \partial_t S - q A_t \\ \partial_x S - q A_x \\ \partial_y S - q A_y \\ 0
    \end{pmatrix},
\end{align}
must satisfy the dispersion relation $k_\mu k^\mu = -m^2$, as well as the Lorentz force law \eqref{eq:Lorentz_force}. The choice of the external electromagnetic field \eqref{eq:Em} gives $k^\nu \nabla_\nu k_z = q F_{z \nu} k^\nu = 0$. Thus, if $\t k{_z}$ vanishes initially, then $k_z = 0$ holds for all time, and the rays that solve the Lorentz force law are constrained to the $xy$-plane.

The transport equation for the intensity \eqref{eq:transp_I} takes the form
\begin{align}
    \t k{^\mu} \t{\nabla\!}{_\mu} \left[\mathcal{I} (t,x,y) \ee^{-z^2/\epsilon} \right]
        = - \mathcal{I} (t,x,y) \ee^{-z^2/\epsilon} \t\nabla{_\mu} \t k{^\mu}\,.
\end{align}
Since $\t k{^z} = 0$, the $z$ dependence of the intensity can be factored out and remains constant over time. Thus, the wave packet will always be localized on the $xy$-plane, and the transport equation for the intensity reduces to
\begin{equation}
    \t k{^\mu} \t{\nabla\!}{_\mu} \mathcal{I} (t,x,y)
        = - \mathcal{I} (t,x,y) \t{\nabla\!}{_\mu} \t k{^\mu}\,.
\end{equation}

Finally, the unit spinor $\psi$ must satisfy the transport equation \eqref{eq:transp_berry}, which is governed by the Berry connection.
Using \cref{eq:Dirac:G metric}, the quantum metric is given by
\begin{align}
    \begin{split}
        \t G{_A_B} = \frac{\t h{_A_B}}{4m^2} \big\{&
            (\t p{^i} \t p{^j} / \mathscr E^2 - \t g{^i^j})
            \t{\dd p}{_i} \otimes \t{\dd p}{_j}
            \\&
            + (\t p{^k} \t p{_k} / \mathscr E)^2
            (\t\p{_i} \ln \lambda \, \t\p{_j} \ln \lambda)
            \t{\dd x}{^i} \otimes \t{\dd x}{^j}
            \\&
            - (\t p{^k} \t p{_k} / \mathscr E^2) (\t p{^i} \t\p{_j} \ln \lambda)
            [
                \t{\dd p}{_i} \otimes \t{\dd x}{^j}
                +
                \t{\dd x}{^j} \otimes \t{\dd p}{_i}
            ]
        \big\}\,,
    \end{split}
\end{align}
where the energy $\mathscr E$ is determined by the dispersion relation
\begin{align}
    \mathscr E
        = - \t p{_0}
        = \sqrt{m^2 + \t g{^i^j} \t p{_i} \t p{_j}}
        = \sqrt{m^2 + \t \delta{^i^j} \t p{_i} \t p{_j} / \lambda^2}
        \,.
\end{align}
Here, the indices $i$, $j$, $k$ range over $1$ and $2$ only.
For a flat space, where $\lambda = 1$, this reduces to the results of Ref.~\cite[Eq. (54)]{Ryu} (up to overall factors that depend on conventions).
However, in the hyperbolic plane, this result shows that the spatial curvature significantly influences the quantum geometry, which has direct implications for the quantum features of hyperbolic quantum materials. For example, the quantum metric plays a central role in the identification of (generalized) momentum-space Landau levels in Chern insulators, which can host fractional quantum Hall states on the lattice \cite{Roy,Thomale,PhysRevB.104.045104, Northe,JieWang,JieWang2}. We envisage a similarly important role for the quantum metric derived here for fractional Chern insulating phases and fractional quantum Hall states on curved space \cite{He,Estienne,Wiegmann,Palumbo2023}, which will be analyzed in future work. Moreover, the study of fermions on hyperbolic surfaces has far-reaching implications since hyperbolic lattices provide platforms for simulating quantum field theories in curved spaces, holographic models \cite{Dey}, and have possible applications in topological quantum computing \cite{Maciejko}.

Finally, our approach can be naturally extended to other quantum systems in any dimension in which the quantum geometric tensor can be affected by the presence of non-trivial background geometry \cite{Jin}.

\section{Conclusions and outlook}

In this paper, we have developed a generalized construction of the quantum geometric tensor by employing the differential-geometric framework of Hermitian vector bundles equipped with connections and sub-bundle projectors. Our approach extends previous formulations by explicitly incorporating an additional curvature contribution arising naturally from the background geometry of constrained quantum states. Furthermore, we showed that the sub-bundle geometry is described by a generalized form of the Gauss–Codazzi–Mainardi equations, in direct analogy with the geometry of submanifolds in Riemannian geometry.

As a concrete illustration, we applied our formalism to Dirac fields in the semiclassical regime, propagating in curved spacetime backgrounds. In this context, we explicitly showed how the additional curvature term arises in the quantum geometric tensor, highlighting its significance for accurately capturing geometric effects in relativistic quantum systems. We considered a specific example of Dirac fermions confined to a hyperbolic plane, which play an important role in hyperbolic Dirac materials. This example elucidates how our generalized quantum geometric tensor encodes subtle geometric features arising from the curvature of the background geometry.

The construction of the quantum geometric tensor described here relies solely on a constant-rank projector on a Hermitian vector bundle with a suitable connection. The geometric nature of our formalism allows for several further extensions by considering the interplay of the shape operators and the quantum geometric tensor with additional geometric structures. For example, a soldering form can induce a notion of torsion for the connection on the vector bundle. Note that this differential-geometric version of torsion is different from the one already used in condensed matter physics \cite{Ahn2,Jankowski4}, where a notion of torsion arises in quantum systems with more than two subspaces. Nevertheless, our results can also be extended to cover these cases, as well as other multi-state geometric objects \cite{mitscherling2024gauge}, by simply considering three or more projectors on the vector bundles.
Additionally, a Levi-Civita connection on the base manifold (describing, for example, spacetime with its Lorentzian metric) can be used to define higher-order quantum geometric tensors obtained by differentiating the shape operators.
Another possibility is to consider more general affine connections on the base manifold, which give rise to non-null torsion and non-metricity. These two tensors, which are independent of the Riemann tensor, have been shown to play an essential role in the study of dislocations \cite{Huang2} and point defects \cite{Palumbo2024} in topological semimetals.

Overall, our results provide a robust and versatile theoretical framework for analyzing quantum geometric properties in curved parameter-dependent quantum systems, particularly those involving non-trivial geometries. This work opens promising avenues for future research into quantum phenomena influenced by curvature effects across diverse areas such as high-energy physics, condensed matter systems, and quantum optics.

\section*{Acknowledgments}

M.A.O.\ is grateful for the hospitality of the Erwin Schrödinger International Institute for Mathematics and Physics, where part of this work was carried out during the Research in Teams program “Weyl Pseudodifferential Calculus with Applications.” This research was funded in whole or in part by the Austrian Science Fund (FWF) \href{https://doi.org/10.55776/PIN9589124}{10.55776/PIN9589124}. Research by T.B.M.\ is funded by the European Union (ERC, GRAVITES, Project No.~101071779). Views and opinions expressed are however those of the authors only and do not necessarily reflect those of the European Union or the European Research Council Executive Agency. Neither the European Union nor the granting authority can be held responsible for them.

\printbibliography
\end{document}